\documentclass[11pt]{article}
\usepackage{vmargin}
\setmarginsrb{.9in}{1in}{.9in}{1in}{0mm}{0mm}{0mm}{7mm}

\usepackage{latexsym}
\usepackage{amssymb}
\usepackage{amsmath}
\usepackage{amsfonts}
\usepackage{amsthm}
\usepackage{txfonts}
\usepackage{url}
\usepackage{stmaryrd}
\usepackage{hyperref}
\usepackage{graphicx}
\usepackage{hyperref}
\usepackage{color}
\usepackage{algorithm}
\usepackage{algorithmic}
\newtheorem{theorem}{Theorem}[section]
\newtheorem{lemma}[theorem]{Lemma}
\newtheorem{corollary}[theorem]{Corollary}

\newtheorem{proposition}[theorem]{Proposition}

\newtheorem{definition}[theorem]{Definition}

\newcommand{\LT}[1]{\mbox{Leaf}(#1)}
\newcommand{\IN}[1]{\mbox{Int}(#1)}

\newcommand{\2}{\vspace{0.2 cm}}
\newcommand{\Al}{\alpha^*}
\newcommand{\Ax}{\alpha}
\newcommand{\Bx}{\beta}
\newcommand{\Mx}[1]{{\bf \mbox{#1}}}
\newcommand{\Cc}{C}
\newcommand{\Rr}{\rho}
\newcommand{\SmallGap}{\vspace{0.2cm}}

\title{Algorithm for Finding $k$-Vertex Out-trees and its Application to $k$-Internal Out-branching Problem}

\author{Nathann Cohen\thanks{INRIA -- Projet MASCOTTE,
2004 route des Lucioles, BP 93
F-06902, Sophia Antipolis Cedex, France,
{\tt nathann.cohen@sophia.inria.fr}} \and
Fedor V. Fomin\thanks{Department of
Informatics, University of Bergen, POB 7803, 5020 Bergen,
Norway, {\tt fedor.fomin|saket.saurabh@ii.uib.no}} \and
Gregory Gutin\thanks{Department of Computer Science, Royal Holloway, University of London,
Egham, Surrey TW20 0EX, UK, {\tt gutin|eunjung|anders@cs.rhul.ac.uk}} \and
\addtocounter{footnote}{-1}
Eun Jung Kim\footnotemark
 \and
\addtocounter{footnote}{-2}
  Saket Saurabh\footnotemark
\and Anders Yeo\footnotemark
}

\begin{document}
\date{}
\maketitle

\begin{abstract}
An out-tree $T$ is  an oriented
tree with only one vertex of in-degree zero.
A vertex $x$ of $T$ is
internal if its out-degree is positive. We design randomized
and deterministic algorithms for deciding whether an  input digraph
contains a given out-tree with $k$ vertices. The algorithms are of
runtime  $O^*(5.704^k)$ and $O^*(5.704^{k(1+o(1))})$, respectively.
We apply the deterministic algorithm to obtain a deterministic
algorithm of runtime $O^*(c^k)$, where $c$ is a constant, for
deciding whether an input digraph contains a spanning out-tree with
at least $k$ internal vertices. This answers in affirmative a
question of Gutin, Razgon and Kim (Proc. AAIM'08).
\end{abstract}

\section{Introduction}

An {\em out-tree} is an oriented tree with only one vertex of
in-degree zero called the {\em root}. The $k$-{\sc Out-Tree} problem is the problem
of deciding for a given parameter $k$, whether an input digraph
contains a given out-tree with $k\ge 2$ vertices.
In their seminal work on
Color Coding Alon, Yuster, and Zwick  \cite{AloYusZwi95}
provided fixed-parameter tractable (FPT) randomized and
deterministic algorithms for $k$-{\sc Out-Tree}. While Alon,
Yuster, and Zwick \cite{AloYusZwi95} only stated that their
algorithms are of runtime $O(2^{O(k)}n)$, however, it is easy to see
(see Appendix),
that their randomized and deterministic algorithms are of complexity\footnote{In this paper we often use the notation
$O^*(f(k))$ instead of $f(k)(kn)^{O(1)}$, i.e., $O^*$  hides not only
constants, but also polynomial coefficients.} $O^*((4e)^k)$ and $O^*(c^k)$,
where $c\ge 4e$.

The main results of \cite{AloYusZwi95}, however, were a new
algorithmic approach called Color Coding and a randomized
$O^*((2e)^{k})$ algorithm for deciding whether a digraph contains
a path with $k$ vertices (the $k$-{\sc Path} problem). Chen et al.
\cite{CHeLuSzeZha07} and Kneis et al. \cite{KneMolRicRos06}
developed a modification of Color Coding, Divide-and-Color, that
allowed them to design a randomized $O^*(4^{k})$-time algorithm
for $k$-{\sc Path}. Divide-and-Color in Kneis et al.
\cite{KneMolRicRos06} (and essentially in Chen et al.
\cite{CHeLuSzeZha07}) is `symmetric', i.e., both colors play
similar role and the probability of coloring each vertex in one of
the colors is 0.5. In this paper, we further develop
Divide-and-Color by making it asymmetric, i.e., the two colors
play different roles and the probability of coloring each vertex in
one of the colors depends on the color.
As a result,  we refine
the result of Alon, Yuster, and Zwick by obtaining
randomized and deterministic algorithms for $k$-{\sc Out-Tree} of
runtime $O^*(5.7^k)$ and $O^*(5.7^{k+o(k)})$ respectively.

It is worth to mention here two recent related results on $k$-{\sc Path}
due to Koutis \cite{Kou08} and Williams \cite{Wil08} based on an algebraic
approach.
 Koutis \cite{Kou08} obtained a randomized
$O^*(2^{3k/2})$-time algorithm for $k$-{\sc Path} and
 Williams \cite{Wil08} extended his ideas resulting in
 a randomized $O^*(2^{k})$-time algorithm for $k$-{\sc
Path}. While the randomized algorithms based on Color Coding and
Divide-and-Color are not difficult to derandomize, it is not the
case for the algorithms of Koutis \cite{Kou08} and Williams
\cite{Wil08}. Thus,  it is unknown whether there are deterministic
algorithms for $k$-{\sc Path} of runtime $O^*(2^{3k/2})$.
Moreover, it is not clear whether the randomized algorithms of Koutis
\cite{Kou08} and Williams \cite{Wil08} can be extended to solve
$k$-{\sc Out-Tree}.

While we believe that the study of fast algorithms for $k$-{\sc Out-Tree}
is a problem interesting on its own, we provide an application of
our deterministic algorithm.
The vertices of an out-tree $T$ of out-degree zero (nonzero) are
{\em leaves} ({\em internal vertices}) of $T$. An {\em
out-branching} of a digraph $D$ is a  spanning subgraph of $D$
which is an out-tree. The {\sc Minimum Leaf} problem is to find an
out-branching with the minimum number of leaves in a given digraph
$D.$ This problem is of interest in database systems
\cite{DemDow00} and the Hamilton path problem is its special case.
Thus, in particular, {\sc Minimum Leaf} is NP-hard. In this paper
we will study the following parameterized version of {\sc Minimum
Leaf} : given a digraph $D$ and a parameter $k$, decide whether $D$
has an out-branching with at least $k$ internal vertices. This
problem denoted {\sc $k$-Int-Out-Branching} was studied for symmetric
digraphs (i.e., undirected graphs) by Prieto and Sloper
\cite{PriSlo03,PriSlo05} and for all digraphs by Gutin et al.
\cite{GutRazKim08}. Gutin et al. \cite{GutRazKim08} obtained an
algorithm of runtime $O^*(2^{O(k\log k)})$ for {\sc $k$-Int-Out-Branching} and
asked whether the problem admits an algorithm of runtime
$O^*(2^{O(k)}).$ Note that no such algorithm has been known even
for the case of symmetric digraphs \cite{PriSlo03,PriSlo05}. In
this paper, we obtain an $O^*(2^{O(k)})$-time algorithm for {\sc $k$-Int-Out-Branching}
using our deterministic algorithm for $k$-{\sc Out-Tree} and an
out-tree generation algorithm.

For a set $X$ of vertices of a subgraph $H$ of a digraph $D$, $N_H^+(X)$ and
$N_H^-(X)$ denote the sets of out-neighbors and in-neighbors of vertices of $X$ in $H$, respectively. Sometimes, when a set has a single element, we will
not distinguish between the set and its element.
In particular, when $H$ is an out-tree and $x$ is a vertex of $H$ which is not its root, the unique in-neighbor of $x$ is denoted by $N_H^-(x)$.
For an out-tree $T$, $\LT{T}$ denotes the set of leaves in $T$ and $\IN{T}=V(T)-\LT{T}$ stands for the set of internal vertices of $T$.

\section{New Algorithms for $k$-{\sc Out-Tree}}\label{TSAsec}

In this section, we introduce and analyze a new randomized algorithm for $k$-{\sc Out-Tree} that uses Divide-and-Color and several other ideas.
We provide an analysis of its complexity and a short discussion of its derandomization.
We omit proofs of several lemmas of this section. The proofs can be found in Appendix.

The following lemma is well known, see \cite{chung}.

\begin{lemma}\label{balanced}
Let $T$ be an undirected tree and let $w:V\rightarrow
\mathbb{R}^+\cup \{0\}$ be a weight function on its vertices. There
exists a vertex $v\in V(T)$ such that the weight of every subtree
$T'$ of $T-v$ is at most $w(T)/2$, where $w(T)=\sum_{v\in
V(T)}w(v)$.
\end{lemma}

Consider a partition $n=n_1+\cdots +n_q,$
where $n$ and all $n_i$ are nonnegative integers and
a bipartition $(A,B)$ of the set $\{1,\ldots ,q\}.$ Let
$d(A,B):=\Big|\sum_{i\in A}n_i-\sum_{i\in B}n_i\Big|.$
Given a set $Q=\{1,\ldots ,q\}$ with a nonnegative integer weight
$n_i$ for each element $i\in Q$, we say that a bipartition $(A, B)$
of $Q$ is {\em greedily optimal} if $d(A,B)$ does not decrease by
moving an element of one partite set into another. The following
procedure describes how to obtain a greedily optimal bipartition in
time $O(q \log{q})$. For simplicity we write $\sum_{i\in A}n_i$ as
$n(A)$.

\begin{algorithm}[h!]
  \caption{\Mx{Bipartition}$(Q,\{n_i:i\in Q\})$}
  \begin{algorithmic}[1]
 \STATE Let $A:=\emptyset$, $B:=Q$.
 \WHILE{$n(A)<n(B)$ and there is an element $i\in B$ with $0<n_i < d(A,B)$}
 \STATE Choose such an element $i\in B$ with a largest $n_i$.
 \STATE $A:=A\cup\{i\}$ and $B:=B-\{i\}.$
 \ENDWHILE
 \STATE Return $(A,B)$.
  \end{algorithmic}
\end{algorithm}

\begin{lemma}\label{largecomp}
Let $Q$ be a set of size $q$ with a nonnegative integer weight $n_i$
for each $i\in Q$. The algorithm \Mx{Bipartition}$(Q,\{n_i:i\in Q\})$
finds a greedily optimal bipartition $A\cup B=Q$ in time
$O(q\log{q})$.
\end{lemma}

This lemma is proved in Appendix.
Now we describe a new randomized algorithm for $k$-{\sc Out-Tree}.

Let $D$ be a digraph and let $T$ be an out-tree on $k$ vertices. Let
us specify a vertex $t\in V(T)$ and a vertex $w\in V(D)$. We call a
copy of $T$ in $D$ a {\em $T$-isomorphic} tree. We say that a
$T$-isomorphic tree $T_{D}$ in $D$ is a $(t,w)$-tree if $w\in
V(T_D)$ plays the role of $t$.

\begin{figure}
   \begin{center}
       \includegraphics [scale =.8, angle =0]{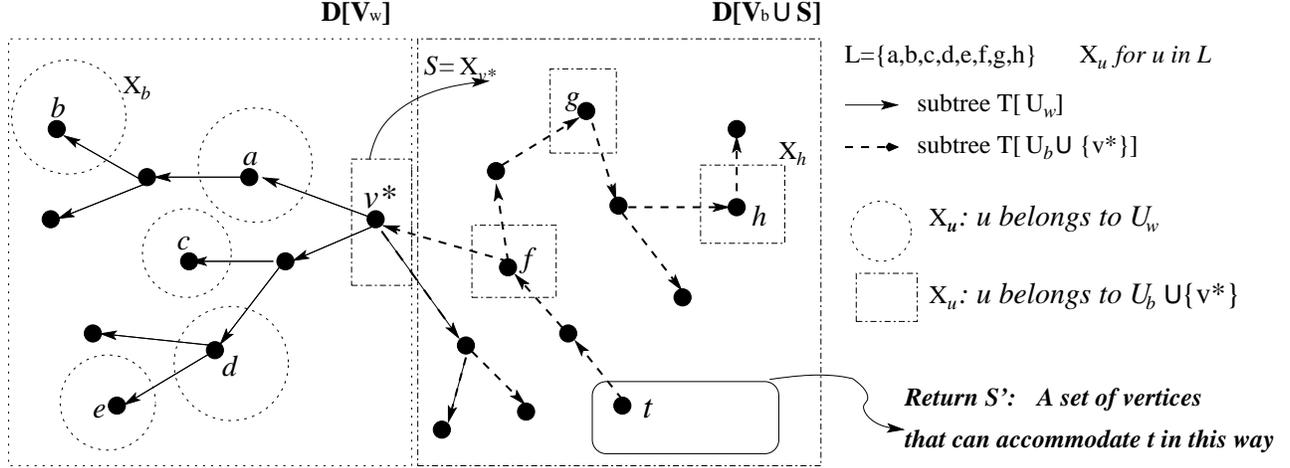}
   \end{center}
   \caption{An example: The given out-tree $T$ is divided into two parts
$T[U_w]$ and $T[U_b\cup \{v^*\}]$ by the splitting vertex $v^*$. The
digraph $D$ contains a copy of $T$ meeting the restrictions on $L$.}
   \label{fig1}
\end{figure}

In the following algorithm \Mx{find-tree}, we have several arguments
other than the natural arguments $T$ and $D$. Two arguments are
vertices $t$ and $v$ of $T$, and the
last argument is a pair consisting of $L\subseteq V(T)$ and
$\{X_u:\ u\in L\}$, where $X_u\subset V(D)$ and $X_u$'s are pairwise
disjoint. The argument $t$ indicates that we want to return, at the
end of the current procedure, the set of vertices $X_t$ such that
there is a $(t,w)$-tree for every $w\in X_t$. The fact that $X_t\neq
\emptyset$ means two points : we have a $T$-isomorphic tree in $D$,
and the information $X_t$ we have can be used to construct a larger
tree which uses the current $T$-isomorphic tree as a building block.
Here, $X_t$ is a kind of `joint'.

The arguments $L\subseteq V(T)$ and $\{X_u:\ u\in L\}$ form a set of
information on the location in $D$ of the vertices playing the role of
$u \in L$ obtained in the way we obtained $X_t$ by a recursive
call of the algorithm.
Let $T_D$ be a
$T$-isomorphic tree; if for every $u\in L$, $T_D$ is a $(v,w)$-tree for some $w\in
X_{u}$ and $V(T_D)\cap X_{u}=\{w\}$, we say that
{\em $T_D$ meets the restrictions on $L$}. The algorithm
\Mx{find-tree} intends to find the set $X_t$ of vertices such that
for every $w\in X_t$, there is a $(t,w)$-tree which meets the
restrictions on $L$;  for illustration, see Figure \ref{fig1}.

The basic strategy is as follows.
We choose a pair $T_A$ and $T_B$
of subtrees of $T$ such that $V(T_A)\cup V(T_B)=V(T)$ and $T_A$ and
$T_B$ share only one vertex, namely $v^*$. We call such $v^*$ a {\em
splitting vertex}.  We call recursively two `\Mx{find-tree}'
procedures on subsets of $V(D)$ to
ensure that the subtrees playing the role of $T_A$ and $T_B$  do not overlap. 
The first call (line 15) tries to find $X_{v^*}$ and the
second one (line 18), using the information $X_{v^*}$ delivered by
the first call, tries to find $X_t$. Here $t$ is a vertex specified
as an input for the algorithm \Mx{find-tree}. In the end, the current
procedure will return $X_t$.

A splitting vertex can
produce several subtrees, but there are many ways to divide them
into two groups ($T_A$ and $T_B$).  To make the algorithm more
efficient, we try to obtain as `balanced' a partition ($T_A$ and
$T_B$) as possible. The algorithm \Mx{tree-Bipartition} is used to
produce a pretty `balanced' bipartition of the subtrees. Moreover we
introduce another argument to have a better complexity behavior. The
argument $v$ is a vertex which indicates whether there is a
predetermined splitting vertex. If $v=\emptyset$, we do not have a
predetermined splitting vertex so we find one in the current
procedure. Otherwise, we use the vertex $v$ as a splitting vertex.

\2

\begin{algorithm}[h!]
  \caption{\Mx{find-tree}($T,D,v,t,L,\{X_u:u\in L\}$), see Figure \ref{fig1}}
  \begin{algorithmic}[1]

\IF{$|V(T)\setminus L|\geq 2$}
    \STATE {\bf for all $u\in V(T)$:} Set $w(u):=0$ if $u\in L$, $w(u):=1$ otherwise.
    \STATE {\bf if $v=\emptyset$ then} Find $v^*\in V(T)$ such that the weight of every subtree $T'$ of $T-v^*$ is at most $w(T)/2$ (see Lemma \ref{balanced}) {\bf else} $v^*:=v$
    \STATE $(WH,BL)$:=tree-Bipartition$(T,t,v^*,L)$.
    \STATE $U_w:=\bigcup_{i\in WH} V(T_i) \cup \{v^*\}$, $U_b:=\bigcup_{i\in BL} V(T_i)$.
    \STATE {\bf for all $u \in L\cap U_w$:} color all vertices of $X_u$ in white.
    \STATE {\bf for all $u \in L\cap (U_b\setminus \{v^*\})$:} color all vertices of $X_u$ in black.
    \STATE $\alpha:=\min\{w(U_w)/w(T),w(U_b)/w(T)\}$.
    \STATE {\bf if $\alpha^2-3\alpha+1\leq 0$ } (i.e., $\alpha \geq (3-\sqrt{5})/2$, see (\ref{par2}) and the definition of $\alpha^*$ afterwards) {\bf then} $v_w:=v_b:=\emptyset$
    \STATE {\bf else if $w(U_w)<w(U_b)$ then} $v_w:=\emptyset$, $v_b:=v^*$ {\bf else} $v_w:=v^*$, $v_b:=\emptyset$.
    \STATE $X_t:=\emptyset$.
    \FOR{$i=1$ to $\left\lceil \frac{2.51}{\Ax{}^{\Ax{}k} (1-\Ax{})^{(1-\Ax{})k}} \right\rceil$}
        \STATE Color the vertices of $V(D)-\bigcup_{u\in L}X_u$ in white or black such that for each vertex the probability to be colored in white is $\alpha$ if $w(U_w)\leq w(U_b)$, and $1-\alpha$ otherwise.
        \STATE Let $V_w$ ($V_b$) be the set of vertices of $D$ colored in white (black).
        \STATE $S:=$find-tree$(T[U_w],D[V_w],v_w,v^*,L\cap U_w,\{X_u:u\in L\cap U_w\})$
        \IF{$S \neq \emptyset$}
        \STATE $X_{v^*}:=S$, $L:=L\cup \{v^*\}$.
        \STATE $S':=$find-tree$(T[U_b\cup \{v^*\}],D[V_b\cup S],v_b,t,(L\cap U_b),\{X_u:u\in (L\cap U_b)\})$.
        \STATE $X_t:=X_t\cup S'$.
        \ENDIF
    \ENDFOR
    \STATE Return $X_t$.

\ELSE [$|V(T)\setminus L| \leq 1$]
    \STATE {\bf if $\{z\}=V(T)\setminus L$ then} $X_z:=V(D)-\bigcup_{u\in L}X_u$, $L:=L\cup \{z\}$.
    \STATE $L^o:=\{$all leaf vertices of $T$\}.
    \WHILE{$L^o\neq L$}
    \STATE Choose a vertex $z\in L\setminus L^o$ s.t. $N^+_T(z)\subseteq L^0$.
    \STATE $X_z:=X_z\cap \bigcap_{u\in N^+_T(z)}N^-(X_u)$; $L^o:=L^o\cup \{z\}$.
    \ENDWHILE
    \RETURN $X_t$
\ENDIF

  \end{algorithmic}
\end{algorithm}

Let $r$ be the root of $T$. To decide whether $D$ contains a copy of $T$, it suffices to run \Mx{find-tree}$(T,D,\emptyset,r,\emptyset,\emptyset)$.

\begin{lemma}\label{argument}
During the performance of find-tree($T,D,\emptyset,r,\emptyset,\emptyset$), the sets $X_u$, $u\in L$ are pairwise disjoint.
\end{lemma}
\begin{proof}
We prove the claim inductively. For the initial call, trivially the sets $X_u$, $u\in L$ are pairwise disjoint since $L=\emptyset$. Suppose that for a call find-tree($T,D,v,t,L,\{X_u:\ u\in L\}$) the sets $X_v$, $v\in L$ are pairwise disjoint. For the first subsequent call in line 15, the sets are obviously pairwise disjoint. Consider the second subsequent call in line 18. If $v^*\in L$ before line 17, the claim is true since $S$ returned by the first subsequent call is contained in $X_{v^*}$. Otherwise, observe that $X_u\subseteq V_b$ for all $u\in L\cap U_b$ and they are pairwise disjoint. Since $X_{v^*}\cap V_b=\emptyset$, the sets $X_u$ for all $u\in L\cap U_b$ together with $X_{v^*}$ are pairwise disjoint.
\end{proof}

\begin{algorithm}[h!]
  \caption{\Mx{tree-Bipartition}$(T,t,v^*,L)$}
  \begin{algorithmic}[1]

 \STATE $T_1,\ldots ,T_q$ are the subtrees of $T-v^*$. $Q:=\{1,\ldots ,q\}$. $w(T_i):=|V(T_i)\setminus L|$, $\forall i\in Q$.
 \IF {$v^*=t$}
 \STATE $(A,B)$:=\Mx{Bipartition}$(Q,\{n_i:=w(T_i):i\in Q\})$
 \STATE {\bf if $w(A)\leq w(B)$ then} $WH:=A$, $BL:=B$. {\bf else} $WH:=B$, $BL:=A$.
 \ELSIF {$t\in V(T_l)$ and $w(T_l)-w(v^*)\geq 0$}
 \STATE $(A,B)$:=\Mx{Bipartition}$(Q,\{n_i:=w(T_i):i\in Q\setminus \{l\}\}\cup \{n_l:=w(T_l)-w(v^*)\})$.
 \STATE {\bf if $l\in B$ then} $WH:=A$, $BL:=B$. {\bf else} $WH:=B$, $BL:=A$.
 \ELSE [$t\in V(T_l)$ and $w(T_l)-w(v^*)< 0$]
 \STATE $(A,B)$:=\Mx{Bipartition}$((Q\setminus \{l\})\cup \{v^*\},\{n_i:=w(T_i):i\in Q\setminus \{l\}\}\cup \{n_{v^*}:=w(v^*)\})$.
 \STATE {\bf if $v^*\in A$ then} $WH:=A-\{v^*\}$, $BL:=B\cup \{l\}$. {\bf else} $WH:=B-\{v^*\}$, $BL:=A\cup \{l\}$.
 \ENDIF
 \RETURN $(WH,BL)$.

  \end{algorithmic}
\end{algorithm}

\begin{lemma}\label{treebip}
Consider the algorithm \Mx{tree-Bipartition} and let $(WH,BL)$ be a bipartition of $\{1,\ldots ,q\}$ obtained at the end of the algorithm. Then the partition $U_w:=\bigcup_{i\in WH} V(T_i) \cup \{v^*\}$ and $U_b:=\bigcup_{i\in BL} V(T_i)$ of $V(T)$ has the the following property.

\noindent 1) If $v^*=t$, moving a component $T_i$ from one partite set to the other does not decrease the difference $d(w(U_w),w(U_b))$.

\noindent 2) If $v^*\neq t$, either exchanging $v^*$ and the component $T_l$ or moving a component $T_i$, $i\neq v^*,l$ from one partite set to the other does not decrease the difference $d(w(U_w),w(U_b))$.
\end{lemma}
\begin{proof}
Let us consider the property 1). The bipartition $(WH,BL)$ is determined in the first `if' statement in line 3 of \Mx{tree-Bipartition}. Then by Lemma \ref{largecomp} the bipartition $(WH,BL)$ is greedily optimal, which is equivalent to the statement of 1).

Let us consider the property 2). First suppose that the bipartition $(WH,BL)$ is determined in the second `if' statement in line 5 of \Mx{tree-Bipartition}. The exchange of $v^*$ and the component $T_l$ amounts to moving the element $l$ in the algorithm \Mx{Bipartition}. Since $(WH,BL)$ is returned by \Mx{Bipartition} and thus is a greedily optimal bipartition of $Q$, any move of an element in one partite set would not decrease the difference $d(WH,BL)$ and the statement of 2) holds in this case.

Secondly suppose that the bipartition $(WH,BL)$ is determined in the third `if' statement in line 8 of \Mx{tree-Bipartition}. In this case we have $w(T_l)=0$ and thus exchanging $T_l$ and $v^*$ and amounts to moving the element $v^*$ in the algorithm \Mx{Bipartition}. By the same argument as above, any move of an element in one partite set would not decrease the difference $d(WH,BL)$ and again the statement of 2) holds.
\end{proof}

Consider the following equation: \begin{equation}\label{par2} \alpha^2-3\alpha +1=0\end{equation} Let $\alpha^*:=(3-\sqrt{5})/2$ be one of its roots. In line 10 of the algorithm \Mx{find-tree}, if $\alpha < \alpha^*$ we decide to pass the present splitting vertex $v^*$ as a splitting vertex to the next recursive call which gets, as an argument, a subtree with greater weight. Lemma \ref{autobalance} justifies this execution. It claims that if $\alpha < \alpha^*$, then in the next recursive call with a subtree of weight $(1-\alpha)w(T)$, we have a more balanced bipartition with $v^*$ as a splitting vertex. Actually, the bipartition in the next step is good enough so as to compensate for the increase in the running time incurred by the biased (`$\alpha < \alpha^*$') bipartition in the present step. We will show this later.

\begin{lemma}\label{autobalance}
Suppose that $v^*$ has been chosen to split $T$ for the present call to \Mx{find-tree}
such that the weight of every subtree of $T-v^*$ is at most $w(T)/2$ and that $w(T) \geq 5$.
Let $\alpha$ be defined as in line $8$ and assume that $\alpha <\alpha^*$.
Let $\{U_1,U_2\}=\{U_w,U_b\}$ such that $w(U_2) \geq w(U_1)$ and let
$\{T_1,T_2\}=\{T[U_w],T[U_b \cup \{v^*\}]\}$ such that $U_1 \subseteq V(T_1)$ and
 $U_2 \subseteq V(T_2)$. Let $\alpha'$ play the role of $\alpha$ in the recursive call using the
tree $T_2$. In this case the following holds:
$\alpha' \geq (1-2\alpha)/(1-\alpha) > \alpha^*.$
\end{lemma}
\begin{proof}
Let $T_1,T_2,U_1,U_2,\alpha,\alpha'$ be defined as in the statement.
Note that $\alpha=w(U_1)/w(T)$.
Let $d=w(U_2)-w(U_1)$ and note that $w(U_1)=(w(T)-d)/2$ and that the following holds
\[ \frac{1-2\alpha}{1-\alpha} = \frac{w(T)-2w(U_1)}{w(T)-w(U_1)} =
\frac{2d}{w(T)+d}.\]

We now consider the following cases.

{\em Case 1. $d=0$:} In this case $\alpha=1/2 > \alpha^*$, a contradiction.

{\em Case 2. $d=1$:} In this case $\alpha^* > \alpha=w(U_1)/(2w(U_1)+1)$, which implies that $w(U_1) \leq 1$. Therefore $w(U_2) \leq 2$ and
$w(T) \leq 3$, a contradiction.

{\em Case 3. $d \geq 2$:} Let $C_1,C_2,\ldots,C_l$ denote the components in $T-v^*$ and without loss of generality assume that
$V(C_1) \cup V(C_2) \cup \cdots \cup  V(C_a) = U_2$ and $V(C_{a+1}) \cup V(C_{a+2}) \cup \cdots \cup  V(C_l) = U_1$.
Note that by Lemma \ref{treebip} we must have $w(C_i) \geq d$ or $w(C_i)=0$ for all $i=1,2,\ldots,l$ except
possibly for one set $C_j$ (containing $t$), which may have
$w(C_j)=1$ (if $w(v^*)=1$).

Let $C_r$ be chosen such that $w(C_r) \geq d$, $1 \leq r \leq a$ and $w(C_r)$ is minimum possible with these constraints.
We first consider the case when $w(C_r) > w(U_2)-w(C_r)$. By the above (and the minimality of $V(C_r)$) we note that
$w(U_2) \leq w(C_r)+1$ (as either $C_j$, which is defined above, or $v^*$ may belong to $V(T_2)$, but not both).
As $w(U_2)=(w(T)+d)/2 \geq  w(T)/2 + 1$ we note that $w(C_r) \geq w(T)/2+d/2 -1$.  As $w(C_r) \leq w(T)/2$ (By the
statement in our theorem) this implies that $d=2$ and $w(C_r)=w(T)/2$ and $w(U_2)=w(C_r)+1$.
If $U_1$ contains at least two distinct components with weight at least $d$ then $w(U_1)>w(U_2)$, a contradiction.
If $U_1$ contains no component of weight at least $d$ then $w(U_1) \leq 1$ and $w(T) \leq 4$, a contradiction.
So $U_1$ contains exactly one component of weight at least $d$. By the minimality of $w(C_r)$ we note that
$w(U_1) \geq w(C_r) = w(U_2)-1$, a contradiction to $d \geq 2$.

Therefore we can assume that $w(C_r) \leq w(U_2)-w(C_r)$, which implies the following
 (the last equality is proved above)
\[
\alpha' \geq \frac{w(C_r)}{w(U_2)} \geq \frac{d}{(w(T)+d)/2} =
\frac{1-2\alpha}{1-\alpha}. \]

As $\alpha<\alpha^*$, we note that $\alpha' \geq (1-2\alpha)/(1-\alpha) > (1-2\alpha^*)/(1-\alpha^*) =\alpha^*$.
\end{proof}

For the selection of the splitting vertex $v^*$ we have two criteria in the algorithm \Mx{find-tree}: (i) {\em `found'} criterion: the vertex is found so that the weight of every subtree $T'$ of $T-v^*$ is at most $w(T)/2$. (ii) {\em `taken-over'} criterion: the vertex is passed on to the present step as the argument $v$ by the previous step of the algorithm. The following statement is an easy consequence of Lemma \ref{autobalance}.

\begin{corollary}\label{twosteppass}
Suppose that $w(T) \geq 5$. If $v^*$ is selected with `taken-over' criterion, then $\alpha > \alpha^*$.
\end{corollary}
\begin{proof}
For the initial call find-tree($T,D,\emptyset,r,\emptyset,\emptyset$) we have $v=\emptyset$ and thus, the splitting vertex $v^*$ is selected with the `found' criterion. We will prove the claim by induction. Consider the first vertex $v^*$ selected with then `taken-over' criterion during the performance of the algorithm. Then in the previous step, the splitting vertex was selected with `found' criterion and thus in the present step we have $\alpha > \alpha^*$ by Lemma \ref{autobalance}.

Now consider a vertex $v^*$ selected with the `taken-over' criterion. Then in the previous step, the splitting vertex was selected with the `found' criterion since otherwise, by the induction hypothesis we have $\alpha > \alpha^*$ in the previous step, and $\emptyset$ has been passed on as the argument $v$ for the present step. This is a contradiction.
\end{proof}

Due to Corollary \ref{twosteppass} the vertex $v^*$ selected in line 3 of the algorithm \Mx{find-tree} functions properly as a splitting vertex. In other words, we have more than one subtree of $T-v^*$ in line 4 with positive weights.

\begin{lemma}\label{split}
If $w(T)\geq 2$, then for each of $U_w$ and $U_b$ found in line 5 of by \Mx{find-tree} we have $w(U_w)>0$ and $w(U_b)>0$.
\end{lemma}
\begin{proof}
For the sake of contradiction suppose that one of $w(U_w)$ and $w(U_b)$ is zero. Let us assume $w(U_w)=0$ and $w(U_b)=w(T)$. If $v^*$ is selected with `found' criteria, each component in $T[U_b]$ has a weight at most $w(T)/2$ and $T[U_b]$ contains at least two components of positive weights. Then we can move one component with a positive weight from $U_b$ to $U_w$ which will reduce the difference $d(U_w,U_b)$, a contradiction. The same argument applies when $w(U_w)=w(T)$ and $w(U_b)=0$.

Consider the case when $v^*$ is selected with ``taken-over"
criteria. There are three possibilities.

{\em Case 1. $w(T)\geq 5$:} In this case we obtain a contradiction
with  Corollary~\ref{twosteppass}.

{\em Case 2. $w(T)=4$:} In the previous step using $T_0$, where
$T\subseteq T_0$, the splitting vertex $v^*$ was selected with
``found" criteria. Then by the argument in the first paragraph, we
have $w(T_0)\geq 5$. A contradiction follows from Lemma~\ref{autobalance}.

{\em Case 3. $2\leq w(T)\leq 3$:} First suppose that $w(v^*)=0$.
Note that $T[U_w]-v^*$ or $T[U_b]$ contains a component of weight
$w(T)$ since otherwise we can move a component with a positive
weight from one partite set to the other and reduce $d(U_w,U_b)$.
Considering the previous step using $T_0$, where $T\subseteq T_0$,
the out-tree $T$ is the larger of $T^0_w$ and $T^0_b$. We pass the
splitting vertex $v^*$ to the larger of the two only when
$\alpha>\alpha^*$. So when $w(T)=3$, we have $3>(1-\alpha^*)w(T_0)$
and thus $w(T^0)\leq 4$, and when $w(T)=2$ we have
$2>(1-\alpha^*)w(T_0)$ and thus $w(T^0)\leq 3$. In either case,
however, $T^0-v^*$ contains a component with a weight greater than
$w(T^0)/2$, contradicting to the choice of $v^*$ in the previous
step (Recall that $v^*$ is selected with `found' criteria in the
previous step using $T^0$).

Secondly suppose that that $w(v^*)=1.$ Then $w(U_w)=w(T)$ and $w(U_b)=0$. We can reduce the difference $d(U_w,U_b)$ by moving the component with a positive weight from $U_w$ to $U_b$, a contradiction.

Therefore for each of $U_w$ and $U_b$ found in line 5 of by \Mx{find-tree} we have $w(U_w)>0$ and $w(U_b)>0$.
\end{proof}

\begin{lemma}\label{correct}
Given a digraph $D$, an out-tree $T$ and a specified vertex $t\in V(T)$, consider the set $X_t$ (in line 22) returned by the algorithm find-tree($T,D,v,t,L,\{X_u:\ u\in L\}$). If $w\in X_t$ then $D$ contains a $(t,w)$-tree that meets the restrictions on $L$. Conversely, if $D$ contains a $(t,w)$-tree for a vertex $w\in V(D)$ that meets the restrictions on $L$, then $X_t$ contains $w$ with probability larger than $1-1/e>0.6321$.
\end{lemma}
\begin{proof}
Lemma \ref{split} guarantees that the splitting vertex $v^*$ selected at any recursive call of \Mx{find-tree} really `splits' the input out-tree $T$ into two nontrivial parts, unless $w(T)\leq 1$.

First we show that if $w\in X_t$ then $D$ contains a $(t,w)$-tree for a vertex $w\in V(D)$ that meets the restrictions on $L$. When $|V(T)\setminus L|\le 1$, using Lemma \ref{argument} it is straightforward to check from the algorithm that the claim holds. Assume that the claim is true for all subsequent calls to \Mx{find-tree}. Since $w\in S'$ for some $S'$ returned by a call in line 18, the subgraph $D[V_b\cup X_{v^*}]$ contains a $T[U_b\cup \{v^*\}]$-isomorphic $(t,w)$-tree $T_D^b$ meeting the restrictions on $(L\cap U_b)\cup \{v^*\}$ by induction hypothesis. Moreover, $X_{v^*}\neq \emptyset$ when $S'\ni w$ is returned and this implies that there is a vertex $u\in X_{v^*}$ such that $T_D^b$ is a $(v^*,u)$-tree. Since $u\in X_{v^*}$, induction hypothesis implies that the subgraph $D[V_w]$ contains a $T[U_w]$-isomorphic $(v^*,u)$-tree, say $T_D^w$.

Consider the subgraph $T_D:=T_D^w \cup T_D^b$. To show that $T_D$ is a $T$-isomorphic $(t,w)$-tree in D, it suffices to show that $V(T_D^w)\cap V(T_D^b)=\{u\}$. Indeed, $V(T_D^w)\subseteq V_w$, $V(T_D^b)\subseteq V_b\cup X_{v^*}$ and $V_w \cap V_b=\emptyset$. Thus if two trees $T_D^w$ and $T_D^b$ share vertices other than $u$, these common vertices should belong to $X_{v^*}$. Since $T_D^b$ meets the restrictions on $(L\cap U_b)\cup \{v^*\}$, we have $X_{v^*}\cap V(T_D^b)=\{u\}$. Hence $u$ is the only vertex that two trees $T_D^w$ and $T_D^b$ have in common. We know that $u$ plays the role of $v^*$ in both trees. Therefore we conclude that $T_D$ is $T$-isomorphic, and since $w$ plays the role of $t$, it is a $(t,w)$-tree. Obviously $T_D$ meets the restrictions on $L$.

Secondly, we shall show that if $D$ contains a $(t,w)$-tree for a vertex $w\in V(D)$ that meets the restrictions on $L$, then $X_t$ contains $w$ with probability larger than $1-1/e>0.6321$. When $|V(T)\setminus L|\le 1$, the algorithm \Mx{find-tree} is deterministic and returns $X_t$ which is exactly the set of all vertices $w$ for which there exists a $(t,w)$-tree meeting the restrictions on $L$. Hence the claim holds for the base case, and we may assume that the claim is true for all subsequent calls to \Mx{find-tree}.

Suppose that there is a $(t,w)$-tree $T_D$ meeting the restrictions on $L$ and that this is a $(v^*,w')$-tree, that is, the vertex $w'$ plays the role of $v^*$. Then the vertices of $T_D$ corresponding to $U_w$, say $T_D^w$, are colored white and those of $T_D$ corresponding to $U_b$, say $T_D^b$, are colored black as intended with probability $\geq  (\alpha^{\alpha}(1-\alpha)^{1-\alpha})^k$. When we hit the right coloring for $T$, the digraph $D[V_w]$ contains the subtree $T_D^w$ of $T_D$ which is $T[U_w]$-isomorphic and which is a $(v^*,w')$-tree. By induction hypothesis, the set $S$ obtained in line 15 contains $w'$ with probability larger than $1-1/e$. Note that $T_D^w$ meets the restrictions on $L\cap U_w$.

If $w'\in S$, the restrictions delivered onto the subsequent call for \Mx{find-tree} in line 17 contains $w'$. Since $T_D$ meets the restrictions on $L$ confined to $U_b-v^*$ and it is a $(v^*,w')$-tree with $w'\in S=X_{v^*}$, the subtree $T_D^b$ of $T_D$ which is $T[U_b\cup \{v^*\}]$-isomorphic meets all the restrictions on $L$. Hence by induction hypothesis, the set $S'$ returned in line 18 contains $w$ with probability larger than $1-1/e$.

The probability $\rho$ that $S'$, returned by \Mx{find-tree} in line 18 at an iteration of the loop, contains $w$ is, thus, $$\rho > (\alpha^{\alpha}(1-\alpha)^{1-\alpha})^k \times (1-1/e)^2 > 0.3995 (\alpha^{\alpha}(1-\alpha)^{1-\alpha})^k.$$ After looping $\lceil(0.3995 (\alpha^{\alpha}(1-\alpha)^{1-\alpha})^k)^{-1}\rceil$ times in line
 12, the probability that $X_t$ contains $w$ is at least $$1-(1-\rho)^{\frac{1}{0.3995 (\alpha^{\alpha}(1-\alpha)^{1-\alpha})^k}}>1-(1-0.3995 (\alpha^{\alpha}(1-\alpha)^{1-\alpha})^k)^{\frac{1}{0.3995 (\alpha^{\alpha}(1-\alpha)^{1-\alpha})^k}}>1-\frac{1}{e}.$$ Observe that the probability $\rho$ does not depend on $\alpha$ and the probability of coloring a vertex white/black.
\end{proof}


\2

The complexity of Algorithm \Mx{find-tree} is analyzed in the following theorem. Its proof given in Appendix is based on Lemmas \ref{split} and \ref{autobalance}.

\begin{theorem}\label{th1}
Algorithm \Mx{find-tree} has running time $O(n^2 k^{\Rr{}}\Cc{}^k)$, where
$w(T)=k$ and $|V(D)|=n$, and $\Cc{}$  and $\Rr{}$ are defined and bounded as
follows:
$$\Cc{} = \left( \frac{1}{\Al{}^{\Al{}} (1-\Al{})^{1-\Al{}}}
\right)^{\frac{1}{\Al{}}},\
\Rr{} = \frac{\ln(1/6)}{\ln(1-\Al{})},\ \Rr{} \leq 3.724, \mbox{ and } \Cc{}\le 5.704.$$
\end{theorem}

Derandomization of the algorithm \Mx{find-tree} can be carried out using the general method presented by Chen et al. \cite{CHeLuSzeZha07} and based on the construction of $(n,k)$-universal sets studied in \cite{naor1995san} (for details, see Appendix). As a result, we obtain the following:

\begin{theorem}
There is a $O(n^2C^{k+o(k)})$ time deterministic algorithm that solves the {\sc $k$-Out-Tree} problem, where $\Cc{}\le 5.704.$
\end{theorem}

\section{Algorithm for {\sc $k$-Int-Out-Branching}}



A {\em $k$-internal out-tree} is an out-tree with at least $k$ internal vertices.
We call a $k$-internal out-tree {\em minimal} if none of its proper subtrees is a $k$-internal out-tree, or
{\em minimal $k$-tree} in short. The {\sc Rooted Minimal $k$-Tree} problem is as follows: given a digraph $D$,
a vertex $u$ of $D$ and a minimal $k$-tree $T$, where $k$ is a parameter, decide whether $D$ contains
an out-tree rooted at $u$ and isomorphic to $T.$
Recall that {\sc $k$-Int-Out-Branching} is the following problem: given a digraph $D$ and a parameter $k$, decide whether
$D$ contains an out-branching with at least $k$ internal vertices. Finally, the {\sc $k$-Int-Out-Tree} problem is stated as follows:
given a digraph $D$ and a parameter $k$, decide whether
$D$ contains an out-tree with at least $k$ internal vertices.

\begin{lemma}\label{minimaltree}
Let $T$ be a
 $k$-internal out-tree. Then  $T$ is minimal if and only if
$|\IN{T}|=k$ and every leaf $u\in \LT{T}$ is the only child of its parent
$N^-(u)$.
\end{lemma}

\begin{proof}
Assume that $T$ is  minimal.
It cannot have more than $k$ internal vertices, because
otherwise by  removing any of its leaves, we obtain
 a subtree of $T$ with at least $k$ internal vertices.
Thus $|\IN{T}|=k$. If there are sibling leaves $u$ and $w$,
then removing one of them provides a subtree of $T$ with
$|\IN{T}|$ internal vertices.

Now, assume that
$|\IN{T}|=k$ and every leaf $u\in \LT{T}$ is the only child of its parent
$N^-(u)$. Observe that every subtree of $T$ can be obtained from $T$ by deleting a leaf of $T$, a leaf in the resulting out-tree, etc.
However, removing any leaf $v$ from $T$ decreases the number of internal vertices,
and thus creates subtrees with at most $k-1$ internal vertices. Thus, $T$ is minimal.

%
\end{proof}

In fact, Lemma \ref{minimaltree} can be used to generate all non-isomorphic minimal $k$-trees.
First, build an (arbitrary) out-tree $T^0$ with $k$ vertices.
Then extend $T^0$ by adding a vertex $x'$ for each leaf $x\in \LT{T^0}$ with an arc $(x,x')$.
The resulting out-tree $T'$ satisfies the properties of Lemma \ref{minimaltree}.
Conversely, by Lemma \ref{minimaltree}, any minimal $k$-tree can be constructed in this way.

\2

{\bf Generating Minimal $k$-Tree (GMT) Procedure}

a. Generate a $k$-vertex out-tree $T^0$ and a set $T':=T^0.$

b. For each leaf $x\in \LT{T'}$, add a new vertex $x'$  and an arc $(x,x')$ to $T'$.

\vspace{2mm}

Due to the following simple observation, to solve {\sc $k$-Int-Out-Tree} for a digraph $D$ it suffices to solve  {\sc Rooted Minimal $k$-Tree} for each vertex $u\in V(D)$ and each minimal $k$-tree $T$ rooted at $u.$

\begin{lemma}\label{containmintree}
Any $k$-internal out-tree rooted at $r$ contains a minimal $k$-tree rooted at $r$ as a subdigraph.
\end{lemma}

Similarly, the next two lemmas show that to solve {\sc $k$-Out-Branching} for a digraph $D$ it suffices to solve  {\sc Rooted Minimal $k$-Tree} for each vertex $u\in S$ and each minimal $k$-tree $T$ rooted at $u,$ where $S$ is the unique strong connectivity component of $D$ without incoming arcs.

\begin{lemma}\cite{BanGut00}
A digraph $D$ has an out-branching rooted at vertex $r\in V(D)$ if and only if $D$ has a unique strong connectivity component $S$ of $D$ without incoming arcs and $r\in S.$ One can check whether $D$ has a unique strong connectivity component and find one, if it exists, in time $O(m+n)$, where $n$ and $m$ are the number of vertices and arcs in $D$, respectively.
\end{lemma}

\begin{lemma}\label{extension}
Suppose a given digraph $D$ with $n$ vertices and $m$ arcs has an out-branching rooted at vertex $r$. Then any minimal $k$-tree rooted at $r$ can be extended to a $k$-internal out-branching rooted at $r$ in time $O(m+n)$.
\end{lemma}
\begin{proof}
Let $T$ be a $k$-internal out-tree rooted at $r$. If $T$ is spanning, there is nothing to prove. Otherwise, choose $u\in V(D)\setminus V(T)$. Since there is an out-branching rooted at $r$, there is a directed path $P$ from $r$ to $u$. This implies that whenever $V(D)\setminus V(T)\neq \emptyset$, there is an arc $(v,w)$ with $v\in V(T)$ and $w\in V(D)\setminus V(T)$. By adding the vertex $w$ and the arc $(v,w)$ to $T$, we obtain a $k$-internal out-tree and the number of vertices $T$ spans is strictly increased by this operation. Using breadth-first search starting at some vertex of $V(T)$, we can extend $T$ into a $k$-internal out-branching in $O(n+m)$ time.
\end{proof}

Since {\sc $k$-Int-Out-Tree} and {\sc $k$-Int-Out-Branching} can be solved similarly, we will only deal with the {\sc $k$-Int-Out-Branching} problem.
We will assume that our input digraph contains a unique strong connectivity component $S$.
Our algorithm called {\em IOBA} for solving {\sc $k$-Int-Out-Branching} for a digraph $D$ runs in two stages.
In the first stage, we generate {\em all} minimal $k$-trees. We use the GMT procedure described above to achieve this.
At the second stage, for each $u\in S$ and
each minimal $k$-tree $T$, we check whether $D$ contains an out-tree rooted at $u$ and isomorphic to $T$ using our algorithm  from the previous section.
We return TRUE if and only if we succeed in finding an out-tree $H$ of $D$ rooted at $u\in S$ which is isomorphic to a minimal $k$-tree.


In the literature, mainly rooted (undirected) trees and not out-trees are studied. However, every rooted tree can be made an out-tree by orienting every edge away from the root and every out-tree can be made a rooted tree by disregarding all orientations. Thus, rooted trees and out-trees are equivalent and we can use results obtained for rooted trees for out-trees.

Otter \cite{Ott48} showed that the number of non-isomorphic
out-trees on $k$ vertices is $t_k=O^*(2.95^k)$. We can generate
all non-isomorphic rooted trees on $k$ vertices using the
algorithm of Beyer and Hedetniemi \cite{BeyHed80} of runtime
$O(t_k)$. Using the GMT procedure we generate all minimal
$k$-trees. We see that the first stage of IOBA can be completed in time
$O^*(2.95^k)$.

In the second stage of IOBA, we try to find a copy of a minimal $k$-tree $T$ in $D$ using our algorithm from the previous section. The running time of our algorithm is $O^*(5.704^k)$. Since the number of vertices of $T$ is bounded from above by $2k-1$, the overall running time for the second stage of the algorithm is  $O^*(2.95^k\cdot 5.704^{2k-1})$. Thus, the overall time complexity of the algorithm is $O^*(2.95^k\cdot 5.704^{2k-1})=O^*(96^k)$.

\vspace{2mm}

We can reduce the complexity with a more refined analysis of the algorithm. The major contribution to the large constant 96 in the above simple analysis comes from the running time of our algorithm from the previous section. There we use the upper bound on the number of vertices in a minimal $k$-tree.  Most of the minimal $k$-trees have less than $k-1$ leaves, which implies that the upper bound $2k-1$ on the order of a minimal $k$-tree is too big for the majority of the minimal $k$-trees.
Let $T(k)$ be the running time of IOBA. Then we have

\begin{equation}\label{eq1}T(k)=O^*\left(\sum_{k+1\leq k'\leq 2k-1} \mbox{(\# of minimal }k-\mbox{trees on }k'\mbox{ vertices) }\times (5.704^{k'})\right)\end{equation}

A minimal $k$-tree $T'$ on $k'$ vertices has $k'-k$ leaves, and thus the out-tree $T^0$ from which $T'$ is constructed has $k$ vertices of which $k'-k$ are leaves. Hence the number of minimal $k$-trees on $k'$ vertices is the same as the number of non-isomorphic out-trees on $k$ vertices with $k'-k$ leaves. Here an interesting counting problem arises.
Let $g(k,l)$ be the number of non-isomorphic out-trees on $k$ vertices with $l$ leaves. Enumerate $g(k,l)$.
To our knowledge, such a function has not been studied yet. Leaving it as a challenging open question, here we give an upper bound on $g(k,l)$ and use it for a better analysis of $T(k)$. In particular we are interested in the case when $l\geq k/2$.

Consider an out-tree $T^0$ on $k\ge 3$ vertices which has $\alpha k$ internal vertices and $(1- \alpha )k$ leaves. We want to obtain an upper bound on the number of such non-isomorphic out-trees $T^0$. Let $T^c$ be the subtree of $T^0$ obtained after deleting all its leaves and suppose that $T^c$ has $\beta k$ leaves. Assume that $\alpha \leq 1/2$ and notice that $\alpha k$ and $\beta k$ are integers. Clearly $\beta < \alpha$.

Each out-tree $T^0$ with $(1- \alpha )k$ leaves can be obtained by appending $(1-\alpha )k$ leaves to $T^c$ so that each of the vertices in $\LT{T^c}$ has at least one leaf appended to it. Imagine that we have $\beta k =|\LT{T^c}|$ and $\alpha k - \beta k=|\IN{T^c}|$ distinct boxes. Then what we are
 looking for is the number of ways to put $(1-\alpha)k$ balls into the boxes so that each of the first $\beta k$ boxes is nonempty. Again this is equivalent to putting $(1-\alpha -\beta)k$ balls into $\alpha k$ distinct boxes. It is an easy exercise to see that this number equals
$
\binom{k-\beta k-1}{\alpha k-1}.
$

Note that the above number does not give the exact value for the non-isomorphic out-trees on $k$ vertices with $(1-\alpha )k$ leaves. This is because we treat an out-tree $T^c$ as a labeled one, which may lead to us to distinguishing two assignments of balls even though the two corresponding out-trees $T^0$'s are isomorphic to each other.

A minimal $k$-tree obtained from $T^0$ has $(1-\alpha)k$ leaves and thus $(2-\alpha)k$ vertices. With the upper bound $O^*(2.95^{\alpha k})$ on the number of $T^c$'s by \cite{Ott48}, by (\ref{eq1}) we have the following:
\begin{align*}
T(k) &= O^*\left(\sum_{\alpha \leq 1/2} \sum_{\beta < \alpha} 2.95^{\alpha k} \binom{k-\beta k-1}{\alpha k-1}(5.704)^{(2-\alpha)k}\right) + O^*\left(\sum_{\alpha >1/2}2.95^{\alpha k}(5.704)^{(2-\alpha)k}\right)\\
 &= O^*\left(\sum_{\alpha \leq 1/2} \sum_{\beta < \alpha} 2.95^{\alpha k} \binom{k}{\alpha k}(5.704)^{(2-\alpha)k}\right)+O^*\left(2.95^k(5.704)^{3k/2}\right)\\
        &= O^*\left(\sum_{\alpha \leq 1/2} \left(2.95^{\alpha} \frac{1}{\alpha ^{\alpha}(1-\alpha)^{1-\alpha}} (5.704)^{(2-\alpha)}\right)^k\right)+O^*(40.2^k)
\end{align*}
The term in the sum over $\alpha \leq 1/2$ above is maximized when $\alpha = \frac{2.95}{2.95+5.704}$, which yields
$T(k)=O^*(49.4^k).$
Thus, we conclude with the following theorem.
\begin{theorem}
{\sc $k$-Int-Out-Branching} is solvable in time $O^*(49.4^k)$.
\end{theorem}

\section{Conclusion}
In this paper we refine the approach of Chen et al. \cite{CHeLuSzeZha07}
and Rossmanith \cite{KneMolRicRos06} based on Divide-and-Color technique.
Our technique is based on a more complicated coloring and within this
technique we refined the result of Alon et al. \cite{AloYusZwi95}
for the $k$-{\sc Out-Tree} problem. It is interesting to see if this
technique can be used to obtain faster algorithms for other parameterized
problems.

As a byproduct of our work, we obtained the first $O^*(2^{O(k)})$
for {\sc $k$-Int-Out-Branching}. We used  the classical result of Otter
\cite{Ott48} that the number of non-isomorphic trees on $k$
vertices is $O^*(2.95^k)$. An interesting combinatorial problem is
to refine this  bound for trees having  $\lfloor \alpha k \rfloor$ leaves for some
$\alpha < 1$.

\section{Appendix}
\subsection{Algorithm of Alon, Yuster and Zwick}\label{Alonsec}

Let $c: V(D)\rightarrow \{1,\ldots ,k\}$ be a vertex $k$-coloring of
a digraph $D$ and let $T$ be a $k$-vertex out-tree contained in $D$
(as a subgraph). Then $V(T)$ and $T$ are {\em colorful} if no pair
of vertices of $T$ are of the same color.

The following algorithm of
\cite{AloYusZwi95} verifies whether $D$ contains a colorful out-tree
$H$ such that $H$ is isomorphic to $T$, when a coloring $c:
V(D)\rightarrow \{1,\ldots ,k\}$ is given. Note that a $k$-vertex
subgraph $H$ will be colorful with a probability of at least $k!/k^k
> e^{-k}$. Thus, we can find a copy of $T$ in $D$ in $e^k$ expected
iterations of the following algorithm.

\begin{algorithm}[h!]
  \caption{$\mathcal{L}(T, r)$}
  \begin{algorithmic}[1]

    \REQUIRE An out-tree $T$ on $k$ vertices, a specified vertex $r$ of $D$
    \ENSURE $\mathcal{C}_T(u)$ for each vertex $u$ of $D$, which is a
family of all color sets that appear on colorful copies of $T$ in $D$,
where $u$ plays the role of $r$
    \medskip
 \IF{$|V(T)|=1$}
 \FORALL{$u\in V(D)$}
 \STATE Insert $\{c(u)\}$ into $\mathcal{C}_T(u)$.
 \ENDFOR
 \STATE Return $\mathcal{C}_T(u)$ for each vertex $u$ of $D$.

 \ELSE
 \STATE Choose an arc $(r',r'')\in A(T)$.
 \STATE Let $T'$ and $T''$ be the subtrees of $T$ obtained by deleting
$(r',r'')$, where $T'$ and $T''$ contains $r'$ and $r''$, respectively.
 \STATE Call $\mathcal{L}(T',r')$.
 \STATE Call $\mathcal{L}(T'',r'')$.
 \FORALL{$u\in V(D)$}
 \STATE Compose the family of color sets $\mathcal{C}_T(u)$ as follows:
 \FORALL{$(u,v)\in A(D)$}
 \FORALL{$ C'\in \mathcal{C}_{T'}(u)$ and $C''\in \mathcal{C}_{T''}(v)$}
 \STATE $C:= C' \cup C''$ if $C'\cap C''=\emptyset$
 \ENDFOR
 \ENDFOR
 \ENDFOR
 \STATE Return $\mathcal{C}_T(u)$ for each vertex $u$ of $D$.
 \ENDIF

  \end{algorithmic}
\end{algorithm}

\begin{theorem}\label{ccrefined}
Let $T$ be a out-tree on $k$ vertices and let $D=(V,A)$ be a
digraph. A subgraph of $D$ isomorphic to $T$, if one exists, can be
found in $O(k(4e)^k\cdot |A|)$ expected time.
\end{theorem}
\begin{proof}
Let $|V(T')|=k'$ and $|V(T'')|=k''$, where $k'+k''=k$. Then
$|\mathcal{C}_{T'}(u)|=\binom{k-1}{k'-1}$ and
$|\mathcal{C}_{T'}(u)|=\binom{k-1}{k''-1}$. Checking $C' \cap
C''=\emptyset$ takes $O(k)$ time. Hence, lines 11-18 requires at most
$\binom{k}{k/2}^2\cdot k|A| \leq k2^{2k}|A|$ operations.

Let $T(k)$ be the number of operations for $\mathcal{L}(T, r)$. We have
the following recursion.

\begin{equation}
T(k)\leq T(k') + T(k'') + k2^{2k-2}|A|
\end{equation}

By induction, it is not difficult to check that $T(k)\leq k4^k|A|$.
\end{proof}

Let $\mathcal{C}$ be a family of vertex $k$-colorings of a digraph
$D$. We call $\mathcal{C}$ an {\em $(n,k)$-family of perfect hashing
functions} if for each $X\subseteq V(D)$, $|X|=k$, there is a
coloring $c\in \mathcal{C}$ such that $X$ is colorful with respect
to $c.$ One can derandomize the above algorithm of Alon et al. by
using any $(n,k)$-family of perfect hashing functions in the obvious
way. The time complexity of the derandomized algorithm depends of
the size of the $(n,k)$-family of perfect hashing functions. Let
$\tau(n,k)$ denote the minimum size of an $(n,k)$-family of perfect
hashing functions. Nilli \cite{nilliCPC3} proved that $\tau(n,k)\ge
\Omega(e^k\log n/\sqrt{k}).$ It is unclear whether  there is
an $(n,k)$-family of perfect hashing functions of size $O^*(e^k)$
\cite{CHeLuSzeZha07}, but even if it does exist, the running time of
the derandomized algorithm would be $O^*((4e)^k).$

\subsection{Proof of Lemma \ref{largecomp}}

\begin{proof}
First we want to show that the values $n_i$ chosen in line 3 of the
algorithm do not increase during the performance of the algorithm.
The values of $n_i$ do not increase because the values of the
difference $d(A,B)$ do not increase during the performance of the
algorithm. In fact, $d(A,B)$ strictly decreases. To see this,
suppose that the element $i$ is selected in the present step. If
$n(A\cup\{i\}) < n(B-\{i\})$, then obviously the difference $d(A,B)$
strictly decreases. Else if $n(A\cup\{i\}) > n(B-\{i\})$, we have
$d(A\cup\{i\},B-\{i\})<n_i<d(A,B)$.

To see that the algorithm returns a greedily optimal bipartition
$(A,B)$, it is enough to observe that for the final bipartition
$(A,B)$, moving any element of $A$ or $B$ does not decrease
$d(A,B)$. Suppose that the last movement of the element $i_0$ makes
$n(A) > n(B)$. Then a simple computation implies that
$d(A,B)<n_{i_0}.$ Since the values of $n_i$ in line 3 of the
algorithm do not increase during the performance of the algorithm,
$n_j\geq n_{i_0}
>d(A,B)$ for every $j\in A$, the movement of any element in $A$
would not decrease $d(A,B)$. On the other hand suppose that
$n(A)<n(B)$. By the definition of the algorithm, for every $j \in B$
with a positive weight we have $n_j\geq d(A,B)$ and thus the
movement of any element in $B$ would not decrease $d(A,B)$. Hence
the current bipartition $(A,B)$ is greedily optimal.

Now let us consider the running time of the algorithm. Sorting the
elements in nondecreasing order of their weights will take
$O(q\log{q})$ time. Moreover, once an element is moved from one
partite set to another, it will not be moved again and we move at
most $q$ elements without duplication during the algorithm. This
gives us the running time of $O(q\log{q})$.
\end{proof}

\subsection{Proof of Theorem \ref{th1}}

\begin{proof}
Let $L(T,D)$ denote the number of times the `if'-statement in line
$1$ of Algorithm \Mx{find-tree} is false (in all recursive calls to
\Mx{find-tree}). We will prove that $L(T,D)\le
R(k)=Bk^{\Rr{}}\Cc{}^k+1$, $B\ge 1$ is a constant whose value will
determined later in the proof. This would imply that the number of
calls to \Mx{find-tree} where the `if'-statement in line $1$ is true
is also bounded by $R(k)$ as if line $1$ is true then we will have
two calls to \Mx{find-tree} (in lines $15$ and $18$). We can
therefore think of the search tree of Algorithm 3 as an out-tree
where all internal nodes have out-degree equal two and therefore the
number of leaves is grater than the number of internal nodes.

Observe that each iteration of the for-loop in line 12 of Algorithm
\Mx{find-tree} makes two recursive calls to \Mx{find-tree} and the time spent in
each iteration of the for-loop is at most $O(n^2)$. As the time
spent in each call of \Mx{find-tree} outside the for-loop is also
bounded by $O(n^2)$ we obtain the desired complexity bound $O(n^2
k^{\Rr{}}\Cc{}^k)$.

Thus, it remains to show that $L(T,D)\le R(k)=Bk^{\Rr{}}\Cc{}^k+1$.
First note that if $k=0$ or $k=1$ then line $1$ is false exactly
once (as there are no recursive calls) and $\min\{R(1),R(0)\}\ge
1=L(T,D)$. If $k\in \{3, 4\}$, then line 1 is false a constant
number of times by Lemma \ref{split} and let $B$ be the minimal
integer such that $L(T,D)\le R(k)=Bk^{\Rr{}}\Cc{}^k+1$ for both
$k=3$ and 4. Thus, we may now assume that $k \geq 5$ and proceed by
induction on $k$.


Let $R'(\Ax{},k)=\frac{6 ((1-\Ax{})k)^{\Rr{}} \Cc{}^{(1-\Ax{})k}
}{\Ax{}^{\Ax{}k} (1-\Ax{})^{(1-\Ax{})k}}.$ Let $\Ax{}$ be defined as
in line $8$ of Algorithm \Mx{find-tree}. We will consider the following two cases
separately.

\2

{\em Case 1, $\Ax{} \geq \Al{}$:} In this case we note that the
following holds as $k \geq 2$ and $(1-\Ax{}) \geq \Ax{}$.

\begin{center}
$\begin{array}{rcl}
\SmallGap{}
 L(T,D) & \leq & \left\lceil \frac{2.51}{\Ax{}^{\Ax{}k} (1-\Ax{})^{(1-\Ax{})k}} \right\rceil
               \times \left( R(\Ax{}k) + R((1-\Ax)k) \right) \\
\SmallGap{}
        & \leq & \frac{3}{\Ax{}^{\Ax{}k} (1-\Ax{})^{(1-\Ax{})k}} \times (2 \cdot R((1-\Ax)k))  \\
        & = &  R'(\Ax{},k).  \\
\end{array}$
\end{center}

By the definition of $\Rr{}$ we observe that $(1-\Al{})^{\Rr{}} =
1/6$, which implies that the following holds by the definition of
$\Cc{}$:

$$ R'(\Al{},k)  =  6 ((1-\Al{})k)^{\Rr{}} \Cc{}^{(1-\Al{})k} \times \Cc{}^{\Al{} k}
            =  k^{\Rr{}} \Cc{}^k
            =  R(k). $$

Observe that

$$\ln(R'(\Ax{},k)) = \ln(6) + \Rr{} \left[ \ln(k) + \ln(1-\Ax{}) \right] +
                  k \left[ (1-\Ax{})\ln(\Cc{}) -\Ax{} \ln(\Ax{}) - (1-\Ax{})\ln(1-\Ax{}) \right] $$

We now differentiate $\ln(R'(\Ax{},k))$ which gives us the
following:

\begin{center}
$\begin{array}{rcl} \frac{\partial(\ln(R'(\Ax{},k)))}{\partial (\Ax{})} & = & \Rr{}
\frac{-1}{1-\Ax{}} +
                 k \left( -\ln(\Cc{}) - ( 1+ \ln(\Ax{})) + (1 + \ln(1-\Ax{}))   \right)  \\
                 & = & \frac{- \Rr{}}{1-\Ax{}} + k \left( \ln \left( \frac{1-\Ax{}}{\Ax{} \Cc{}} \right) \right). \\
\end{array}$
\end{center}

Since $k \geq 0$  we note that the above equality implies that
$R'(\Ax{},k)$ is a decreasing function in $\Ax{}$ in the interval
$\Al{} \leq \Ax{} \leq 1/2$. Therefore $L(T,D) \leq R'(\Ax{},k) \leq
R'(\Al{},k) = R(k)$, which proves Case 1.

{\em Case 2, $\Ax{} < \Al{}$:} In this case we will specify the
splitting vertex when we make recursive calls using the larger of
$U_w$ and $U_b$ (defined in line $5$ of Algorithm \Mx{find-tree}). Let $\Ax{}'$
denote the $\Ax{}$-value in such a recursive call.
  By Lemma \ref{autobalance} we note that the following holds :

$$ \frac{1}{2} \geq \Ax{}' \geq \frac{1-2\Ax{}}{1-\Ax{}} > \Al{}. $$

  Analogously to Case 1 (as $R'(\Ax{}',(1-\Ax{})k)$ is a decreasing function in $\Ax{}'$ when $1/2 \geq \Ax{}' \geq \Al{}$)
 we note that the
$L$-values for these recursive calls are bounded by the following,
where $\Bx{} = \frac{1-2\Ax{}}{1-\Ax{}}$ (which implies that
$(1-\Ax{})(1-\Bx{})= \Ax{}$):

\begin{center}
$\begin{array}{rcl}
\SmallGap{}
R'(\Ax{}',(1-\Ax{})k) & \leq &  R'\left( \Bx{}, (1-\Ax{})k \right) \\
\SmallGap{}
            &  =  &  \frac{3}{ \left( \Bx{}^{\Bx{}} (1-\Bx{})^{(1-\Bx{})} \right)^{(1-\Ax{})k}} \times 2 \times R((1-\Bx)(1-\Ax{})k)  \\
            &  =  &  \frac{6 R(\Ax{}k)}{ \left( \Bx{}^{\Bx{}} (1-\Bx{})^{(1-\Bx{})} \right)^{(1-\Ax{})k} }.  \\
\end{array}$
\end{center}

Thus, in the worst case we may assume that
$\Ax{}'=\Bx{}=(1-2\Ax{})/(1-\Ax{})$ in all the recursive calls using
the larger of $U_w$ and $U_b$. The following now holds (as $k \geq
2$).

\begin{center}
$\begin{array}{rcl}
\SmallGap{}
 L(T,D) & \leq & \left\lceil \frac{2.51}{\Ax{}^{\Ax{}k} (1-\Ax{})^{(1-\Ax{})k}} \right\rceil
               \times \left( R(\Ax{}k) + R'(\Ax{}',(1-\Ax{})k) \right) \\
\SmallGap{}
        & \leq & \frac{3}{\Ax{}^{\Ax{}k} (1-\Ax{})^{(1-\Ax{})k}} \times R(\Ax{}k) \times
                 \left( 1+ \frac{6}{ \left( \Bx{}^{\Bx{}} (1-\Bx{})^{(1-\Bx{})} \right)^{(1-\Ax{})k} }  \right) \\
        & \leq & \frac{3R(\Ax{}k) }{\Ax{}^{\Ax{}k} (1-\Ax{})^{(1-\Ax{})k}} \times
                  \frac{7}{ \left( \Bx{}^{\Bx{}} (1-\Bx{})^{(1-\Bx{})} \right)^{(1-\Ax{})k} } \\
\end{array}$
\end{center}

  Let $R^*(\Ax{},k)$ denote the bottom right-hand side of the above equality (for any value of $\Ax{}$).
By the definition of $\Rr{}$ we note that $ \Rr{} = \frac{2
\ln(1/6)}{ 2 \ln(1-\Al{})} = \frac{\ln(1/36)}{\ln(\Al)}$, which
implies that $(\Al{})^{\Rr{}}= 1/36$. By the definition of $\Cc{}$
and the fact that if $\Ax{}=\Al{}$ then
$\Bx{}=(1-2\Al{})/(1-\Al{})=\Al{}$, we obtain the following:

\begin{center}
$\begin{array}{rcl}
\SmallGap{}
 R^*(\Al{},k) & = &  \frac{3 R(\Al{}k) }{\Al{}^{\Al{}k} (1-\Al{})^{(1-\Al{})k}} \times
                  \frac{7}{ \left( \Al{}^{\Al{}} (1-\Al{})^{(1-\Al{})} \right)^{(1-\Al{})k} }  \\
\SmallGap{}
              & = &  21 \cdot R(\Al{}k) \cdot \Cc{}^{\Al{}k} \cdot \Cc{}^{\Al{} (1-\Al{}) k} \\
\SmallGap{}
              & = &  21 \Al{}^{\Rr{}} k^{\Rr{}} \Cc{}^{\Al{}k} \times \Cc{}^{(2\Al{}-\Al{}^2)k} \\
\SmallGap{}
              & = &  21 \Al{}^{\Rr{}} R(k) \\
              & < &  R(k). \\
\end{array}$
\end{center}

  We will now simplify $R^*(\Ax{},k)$ further, before we differentiate $\ln(R^*(\Ax{},k))$.
Note that $\Bx{} = \frac{1-2\Ax{}}{1-\Ax{}}$ implies that $(1-\Ax{})(1-\Bx{})= \Ax{}$ and
$\Bx(1-\Ax{})=1-2\Ax{}$.

\begin{center}
$\begin{array}{rcl}
\SmallGap{}
 R^*(\Ax{},k) & = &   \frac{21 R(\Ax{}k) }{\Ax{}^{\Ax{}k} (1-\Ax{})^{(1-\Ax{})k}} \times
                  \frac{1}{ \left( \Bx{}^{\Bx{}} (1-\Bx{})^{(1-\Bx{})} \right)^{(1-\Ax{})k} }   \\
\SmallGap{}
              & = &   \frac{21 (\Ax{}k)^{\Rr{}} \Cc^{\Ax{}k}  }{\Ax{}^{\Ax{}k} (1-\Ax{})^{(1-\Ax{})k}} \times
                  \frac{1}{ \left( \frac{1-2\Ax{}}{1-\Ax{}} \right)^{(1-2\Ax{})k}
                                        \left( \frac{\Ax{}}{1-\Ax{}} \right)^{\Ax{}k} }  \\
              & = &  21 (\Ax{}k)^{\Rr{}}  \left( \frac{ \Cc^{\Ax{}} }{\Ax{}^{2\Ax{}} (1-2\Ax{})^{(1-2\Ax{})}} \right)^k. \\
\end{array}$
\end{center}

Thus, we have the following:

$$\ln(R^*(\Ax{},k)) = \ln(21) + \Rr{} \left( \ln(k) + \ln(\Ax{}) \right) +
                  k \left( \Ax{}\ln(\Cc{}) -2\Ax{} \ln(\Ax{}) - (1-2\Ax{})\ln(1-2\Ax{}) \right). $$

We now differentiate $\ln(R^*(\Ax{},k))$ which gives us the
following:

\begin{center}
$\begin{array}{rcl}
\SmallGap{}
\frac{\partial(\ln(R^*(\Ax{},k)))}{\partial(\Ax{})} & = & \frac{\Rr{}}{\Ax{}} + k \left( \ln(\Cc{}) -2(1+\ln(\Ax{})) + 2(1+\ln(1-2\Ax{})) \right) \\
  & = & \frac{\Rr{}}{\Ax{}} + k \left( \ln\left( \frac{\Cc{} (1-2\Ax{})^2}{\Ax{}^2} \right) \right) \\
\end{array}$
\end{center}

Since  $k \geq 0$  we note that the above equality implies that
$R^*(\Ax{},k)$ is an increasing function in $\Ax{}$ in the interval
$1/3 \leq \Ax{} \leq \Al{}$. Therefore $L(T,D) \leq R^*(\Ax{},k)
\leq R^*(\Al{},k) < R(k)$, which proves Case 2.
\end{proof}

\subsection{Derandomization of Our Randomized Algorithm for $k$-{\sc Out-Tree}}\label{Derandomdsec}
In this subsection we discuss the derandomization of the algorithm \Mx{find-tree} using the general method presented by Chen et al. \cite{CHeLuSzeZha07} and based on the construction of $(n,k)$-universal sets studied in \cite{naor1995san}.
\begin{definition}
An $(n,k)$-universal set $\mathcal F$ is a set of functions from $[n]$ to $\{0,1\}$, such that for every subset $S\subseteq [n],|S|=k$ the set $\mathcal F|_{S}=\{f|_S,f\in T\}$ is equal to the set $2^S$ of all the functions from $S$ to $\{0,1\}$.
\end{definition}
\begin{proposition}[\cite{naor1995san}]
\label{propuniversalsets}
There is a deterministic algorithm of running time $O(2^kk^{O(\log k)} n\log n)$ that constructs an $(n,k)$-universal set $\mathcal F$ such that $|\mathcal F|=2^kk^{O(\log k)}\log n$
\end{proposition}

We explain how Proposition \ref{propuniversalsets} is used to achieve a deterministic algorithm for the {\sc $k$-Out-Tree} problem. Let $V(G')=\{v_1,\dots,v_n\}$. First, we construct an $(n,k)$-universal set $\mathcal F$ of size $2^kk^{O(\log k)}\log n$ (this can be done in time $O(2^kk^{O(\log k)}n\log n)$). Then we call the algorithm \Mx{find-tree} but replace steps 13 and 14 by the following steps:
\begin{enumerate}
\item[13] {\bf for} each function $f\in \mathcal F$ {\bf do}
\item[14] $\forall i$ such that $x_i\in V(D)-\bigcup_{u\in L}X_u$, let $v_i$ be colored in white if $f(i)=0$ and in black otherwise
\end{enumerate}
Note that this replacement makes the algorithm {\bf find-tree} become deterministic. Then, since $\mathcal F$ is a $(n,k)$-universal set and if there is a subgraph isomorphic to $T$ in $D$, there is a function in $\mathcal F$ such that the vertices corresponding to $U_w$ in $D$ with be colored in white while the vertices corresponding to $U_b$ will be colored in black. Using induction on $k$, we can prove that this deterministic algorithm correctly returns the required tree at the condition that such a tree exists in the graph. We can also derive the running time of this deterministic algorithm to find a complexity of $O(n^2C^{k+o(k)})$.

\begin{theorem}
There is a $O(n^2C^{k+o(k)})$ time deterministic algorithm that solves the {\sc $k$-Out-Tree} problem.
\end{theorem}

%
%
%
%
%
%

\end{document}